\input{aipcheck}

\documentclass[
    ,final            
,numberedheadings 
  ]
  {aipproc}

\layoutstyle{8x11double}

\begin{document}

\title{Overview of MHz air shower radio experiments and results}



\author{Beno\^it Revenu}{
  address={\'Ecole des Mines de Nantes - Universit\'e de Nantes - CNRS/IN2P3\\4 rue Alfred Kastler, BP20722, 44307 Nantes C\'EDEX 03, FRANCE}
}



\begin{abstract}
In this paper, I present a review of the main results obtained in the last 10 years in the field of radio-detection of cosmic-ray air showers in the MHz range. All results from all experiments cannot be reported here so that I will focus on the results more than on the experiments themselves. Modern experiments started in 2003 with CODALEMA and LOPES. In 2006, small-size autonomous prototypes setup were installed at the Pierre Auger Observatory site, to help the design of the Auger Engineering Radio Array (AERA). We will discuss the principal aspects of the radio data analysis and the determination of the primary cosmic ray characteristics: the arrival direction, the lateral distribution of the electric field, the correlation with the primary energy, the emission mechanisms and the sensitivity to the composition of the cosmic rays.
\end{abstract}

\maketitle


\section{Introduction}

In 1971, Allan~\cite{allan1971} gave a review of the latest results, at that time, concerning the detection of extensive air showers initiated by high-energy cosmic rays.
The initial motivations for studying the radio signal was the possibility to build an array of widely spaced receivers to detect extensive air showers at the highest energies (above 10~EeV), in a cheap way. It was also proposed to use the radio signal in conjonction with a particle array to get additional information on the longitudinal profile, providing valuable constraints on the nature of the primary cosmic ray. Using the data available at that time, Allan concluded that the main signal should be polarized in the direction of the vector $\mathbf{v}\times\mathbf{B}$, where $\mathbf{v}$ is the direction of the shower axis and $\mathbf{B}$ the direction of the geomagnetic field. The mechanism responsible for this specific polarization is the geomagnetic contribution due to the Lorentz force acting on each secondary charged particles, in particular the electrons and positrons. It was predicted that the charge excess contribution should be less important but still detectable, mainly for incoming directions close to the direction of the geomagnetic field. It is also stated that the electric field amplitude extrapolated on the shower axis, over the range 32-55~MHz, is proportional to the primary energy. Finally, for a given event observed by several detectors, the electric field amplitude decreases exponentially with the axis distance. The equation proposed to fully describe the electric field $\varepsilon_\nu$ observed by a single detector at a distance $R$ of the shower axis is:
\begin{equation}
\varepsilon_\nu=20\left(\frac{E_P}{10^{17}~\mathrm{eV}}\right) \sin\alpha \cos\theta \exp\left(-\frac{R}{R_0(\nu,\theta)}\right)\label{profile}
\end{equation}
in $\mu\mathrm{V}~\mathrm{m}^{-1}~\mathrm{MHz}^{-1}$, where $E_P$ is the primary energy, $\alpha$ the angle between the shower axis and the geomagnetic field, $\theta$ the zenith angle, $R$ the distance between the detector and the shower axis and $R_0$ the attenuation length of the electric field. The lateral distribution function (LDF) proposed by Allan depends on the axis distance leading to an azimuthal invariance of the electric field with respect to the shower axis.
This formula was used as a starting point at the beginning of the years 2000 for the design of the CODALEMA~\cite{coda2006} and LOPES~\cite{lopes1} experiments. These first modern experiments were triggered by an array of particle detectors  in order to search a posteriori for a radio counterpart to a shower candidate. The next challenge consisted in setting up fully autonomous and independent radio detectors to test the possibility to build a full stand-alone radio array. A first type of such radio array was installed in 2006 at the center of the surface detector (SD) of the Pierre Auger Observatory, in the scope of the Auger Engineering Radio Array (AERA), experiment which finishes its phase 1 in 2012. AERA benefits from the SD and FD (fluorescence detector) reconstructions, providing high-quality super-hybrid events. The LOFAR~\cite{Corstanje:2011rx} experiment also detects the radio emission of air showers. The complementary particle detector, LORA~\cite{Thoudam:2011zs}, helps in triggering and identifying the cosmic rays detected by LOFAR.

We present in this overview the updated results of the field. We will discuss the lateral distribution function, the correlation with the primary energy and we will insist on the current situation concerning the emission mechanisms that can be determined through the polarization of the total electric field.



\section{Detection of the electric field emitted by EAS}
The secondary electrons and positrons created during the development of the air shower form a pancake-shape particle front moving at the speed of light. The thickness of the shower front is of the order of 1~m on the shower axis, up to $\sim 10$~m far from the axis. These particles suffer a systematic opposite drift caused by the Lorentz force due to the Earth's magnetic field generating a coherent emission of electromagnetic waves in the 1-500 MHz range. The amplitude of the resulting macroscopic current and its variation depends on the number of charges particles of the shower and therefore, can provide information on the longitudinal profile. The electric field produced this way is expected to have a polarization following the direction of the Lorentz force, $\mathbf{v}\times\mathbf{B}$. This mechanism is known as the geomagnetic contribution as discussed in the introduction. Another mechanism is due to the variation of the excess of electrons due to the annihilation of positrons and knock-on electrons, known as the charge-excess contribution. This corresponds to the Askaryan~\cite{ask1962} effect in the air, leading to a radially-polarized electric field, also in the MHz domain. It is therefore possible to disentangle these two contributions through the different polarizations of the associated electric fields. Various simulation codes are available, a review is given in this conference, see~\cite{huege_arena2012}. This electric field is usually detected by systems composed of antennas, precise time tagging system (GPS for instance) and the signal is digitized by fast ADCs in order to be able to compute the Fourier spectrum up to some hundreds of MHz. The electric field is filtered in a frequency band above the AM ($\sim 20$~MHz) and below the FM (80~MHz). Following the antenna used, the filtering can be done in more restricted bands accordingly to the frequency response of the device used. In general, the electric field is measured in, at least, two horizontal orthogonal polarizations; this permits to reconstruct the polarization angle in the horizontal plane and to check for the underlying emission mechanism.

The two first modern radio experiments, CODALEMA and LOPES, led to many progresses in the understanding of the emission of the electric field by EAS. These experiments detect the EAS initiated by cosmic rays with an energy between $10^{16}$ and $10^{18}$~eV and follow the same principle: they use a particle array which 
triggers the radio array. The particle array for CODALEMA is an array of 17 scintillators and in the case of LOPES, the particle array is the KASCADE-Grande experiment. 
LOPES is installed close to the city of Karlsruhe which is an electromagnetically noisy environment. The solution adopted by LOPES is to use interferometric technics, using as an input the high-quality information provided by KASCADE-Grande. A digital beam-forming is applied to inter-phase the time series of all LOPES antennas using the arrival direction given by KASCADE-Grande. Then, a cross-calibration procedure is used to estimate the global amplitude of the electric field. For high signal-to-noise ratio events, the electric field can be estimated for each antenna, this permits to compute the LDF of these events.
On the contrary, the CODALEMA experiment is installed in the radio observatory of Nan\c{c}ay which is a protected area in the sense that the electromagnetic emissions of the neighbourhood is controlled. This permits to use the data of the radio array of CODALEMA independently of the data of the particle array, once the EAS is clearly identified and validated by the particle array. Each radio detector provides the electric field as a function of time with a high sampling rate (typically 500~Ms/s or 1~Gs/s) so that it's possible to measure the time of the maximum of the signal.

\section{Arrival direction}
From the measurements of the electric field by several detectors, it is possible, in the very same way than with a particle array, to reconstruct the incoming direction of the EAS. Provided the ground coordinates of at least three non-aligned detectors and the time of transit of the electromagnetic wave of the shower in these detectors has been measured, we can triangulate to estimate $\theta$ and $\phi$, the zenith and azimuth angles of the shower, respectively. For higher multiplicity events, we can compute the radius of curvature of the electromagnetic wave assuming a spherical front. The time resolution of GPS receivers allow to reach an angular resolution of the order of a fraction of degree, even for low-multiplicity events. Artificial sources are commonly used to time-calibrate the radio detectors. For instance, the detection of airplane transients represents a gold mine because it's possible to inter-calibrate the detectors, to study the antenna lobe sensitivity and also to compute the angular resolution. It has been proven that this angular resolution can reach $0.5^\circ$, even for a small number of detectors spaced by 140~m only~\cite{revenuicrc2011}. Octocopter flights have been used for instance in LOFAR, LOPES and AERA. The arrival direction estimation is important at two levels: it permits to identify --~and suppress~-- anthropic events (mainly coming from the horizon, apart for airplanes) and to identify cosmic ray events by a comparison with the arrival direction given by the particle array. If the directions agree within a certain angle (typically less than $20^\circ$) and the time of the shower is the same within a certain time window (less than $100$~ns for instance), then the event is seen by both arrays.

\section{Lateral distribution function}
\begin{center}
\begin{figure}[!ht]
\includegraphics[scale=0.4,draft=false]{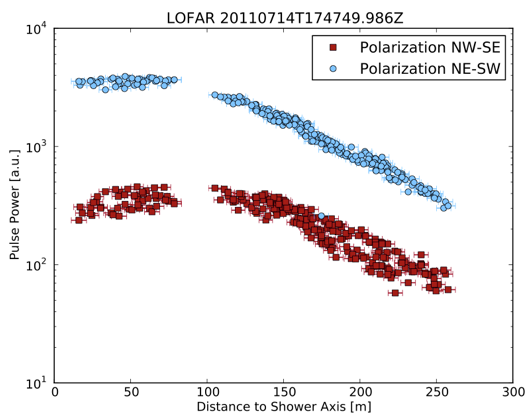}
\caption{Air shower detected by the LORA/LOFAR experiments. The flattening of the profile is interpreted as an effect of the air refractive index leading to a Cerenkov ring close (less than $\sim 100$~m) to the shower axis.}\label{lofarevt}
\end{figure}
\end{center}

The LDF describes the electric field amplitude as a function of the distance to the shower axis. The starting point for the function describing the LDF is an exponential, as first proposed by Allan. Most of the radio-detected events are well described by an exponential profile (see Eq.~\ref{profile}). Nevertheless, a non-negligible fraction of events (of the order of 20\%) present a flattening for detectors close to the shower axis, in particular for inclined showers, as reported by the LOPES collaboration in~\cite{lopes2009ci}. The same observation holds for the CODALEMA data and the LOFAR collaborations. A very nice example of such event is given in~\cite{Corstanje:2011rx}. This event was detected by 5 LOFAR stations corresponding to more than 200 independent measurements of the electric field at different axis distances; the radio pulse power as a function of axis distance is presented in Figure~\ref{lofarevt}. The flattening of the profile is very clear for antennas located at less than 100~m from the shower axis.
The flattening of the electric field profiles can be understood as the effect of the air refractive index. At the sea level, $n\sim 1.0003$ and this value decreases with increasing altitude; the consideration of realistic values of the air refractive index in the simulations permits to reproduce events with a flat profile close to the shower axis which is the consequence of a Cerenkov ring (see~\cite{huege_arena2012}, \cite{coreas_2012}, \cite{marin_arena2012}, \cite{carvalho_arena2012}). One should be cautious when using the usual 1D exponential profile, in particular when considering the electric field measurements close to the shower axis. Moreover, the electric field should not exhibit azimuthal invariance with respect to the shower axis. This property is predicted by all the simulation codes and has also been observed in the CODALEMA data~\cite{marinicrc2011}. More complicated 2D LDFs should be used in the future analyses but at the time of this conference, no analysis using 2D LDF has been presented. The minimization of the $\chi^2$ based on the model $s(d)=s_0 \exp(-d/d_0)$ gives the core position through the axis distance $d=|\mathbf{n}\times\mathbf{CA}|$ where $\mathbf{n}$ is the shower axis and $\mathbf{CA}$ is the vector between the core position and the position of antenna $A$. The minimization also provides the attenuation length $d_0$ and the on-axis signal $s_0$. The on-axis signal can be defined as the actual electric field if the amplitudes $s(d)$ have properly been deconvoluted for the antenna response (arrival direction and frequency gain). This deconvolution is mandatory in order to be able to study the correlation between electric field and energy of the primary cosmic ray. Using simulations, it has been demonstrated that the shower-to-shower fluctuations are minimized when considering the electric field at a distance of $\sim 110$~m of the shower axis. It could be therefore interesting to consider this value instead of the on-axis electric field \cite{palmieri2012}.

\section{Correlation with the primary energy}
\begin{center}
\begin{figure}[!ht]
\includegraphics[scale=0.3,width=7.5cm,draft=false]{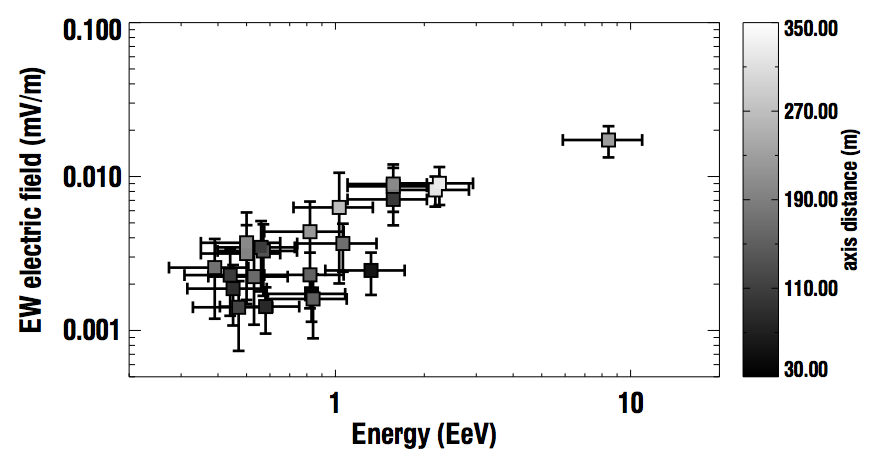}
\caption{Correlation of the extrapolated on-axis electric field with the primary energy estimated by the Auger SD. The Pearson correlation coefficient is $0.81^{+0.12}_{-0.46}$ at 95\% CL, taking into account all sort of systematic and statistical errors.}\label{raugercorr}
\end{figure}
\end{center}

Hybrid arrays usually use the particle detector array for the energy estimation. As stated before, the LOPES experiment relies on the KASCADE-Grande reconstruction, the CODALEMA experiment uses the scintillator array and the AERA experiment depends on the Auger SD and/or FD reconstruction. Allan proposed a linear relation between the on-axis electric field and the primary energy.
This relation is confirmed by the LOPES 30 data, using the pulse height in the east-west (EW) polarization: $\varepsilon_\mathrm{EW}\propto E_p^{0.95\pm 0.04}$~(see \cite{2008ICRC....4...83H}).
More recently, the AERA experiment reported the correlation between the electric field and the primary energy determined by the Auger SD and FD. The signal is recorded in the three polarizations EW, north-south (NS) and vertical (V). The AERA radio detectors have been fully calibrated and the measurements are deconvoluted to estimate the 3D electric field vector~\cite{Abreu:2012pi}. Then the Hilbert envelope is computed for the three directions and the total signal strength is defined as the maximum of the 3D Hilbert envelope~\cite{glaser_arena2012}. The electric field is finally interpolated at 110~m of the shower axis, where the energy resolution is maximum, and the correlation with the primary energy is compatible with a linear relation. The AERA energy correlation plot will be published in a forthcoming paper. The Pierre Auger collaboration also reported a positive correlation at 99.99\% CL between the on-axis electric field and the primary energy, obtained with the (pre-AERA) RAuger setup, as can be seen in Figure~\ref{raugercorr} extracted from~\cite{1748-0221-7-11-P11023}.
It has been reported during this conference the quasi-linear correlation of the CODALEMA data with the primary energy using raw data (not deconvoluted for the antenna response)~\cite{lautridou_arena2012}. Due to the different estimators used in the community (either the on-axis electric field or its value at $\sim 110$~m), one should be cautious when comparing the slopes of the correlations reported by several experiments.

\section{Emission mechanisms and polarization}
\begin{center}
\begin{figure}[!ht]
\includegraphics[scale=0.35,draft=false]{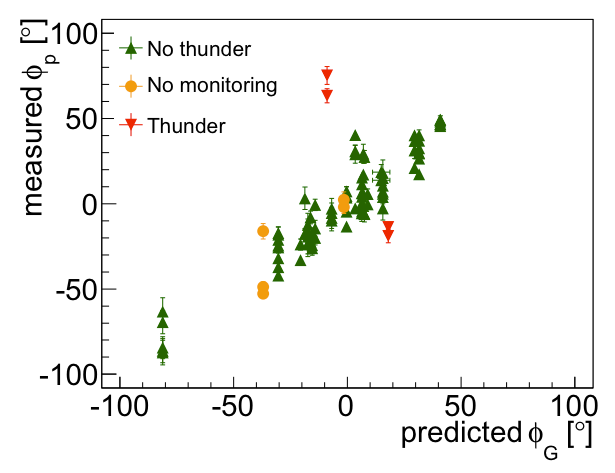}
\caption{Polarization angle in the horizontal plane of the detected electric field as a function of the expected polarization angle in case of a pure geomagnetic contribution. Green triangles correspond to regular atmospheric conditions. Yellow circles correspond to events for which the atmospheric monitoring was not working. Red triangles corresponds to events detected during thunderstorms.}\label{figpolarangle}
\end{figure}
\end{center}

\begin{center}
\begin{figure}[!ht]
\includegraphics[scale=0.35,draft=false]{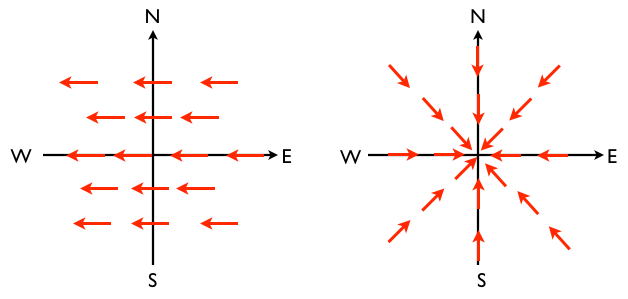}
\caption{Polarization pattern at the ground level for a vertical shower in a geomagnetic field oriented along the NS axis. The left figure corresponds to the geomagnetic contribution and the right figure, to the charge-excess contribution.}\label{polscheme}
\end{figure}
\end{center}

The geomagnetic contribution to the total electric field is dominant, as stated in the 1960s and confirmed with much more statistics and better data some years ago by CODALEMA~\cite{coda2009}, \cite{revenuicrc2009} and LOPES~\cite{2008ICRC....4...83H}. In the southern hemisphere, the RAuger prototype reported that the arrival directions of the detected air showers were in good agreement with a $\mathbf{v}\times\mathbf{B}$ effect. Using the AERA data, we studied the correlation in the horizontal plane of the expected polarization angle $\phi_G$ of the geomagnetically-induced electric field with the measured polarization angle $\phi_P$. The correlation is excellent, as can be seen in Figure~\ref{figpolarangle} (extracted from~\cite{schoorlemmerICTAPP2011}).
As discussed in the introduction, the electrons in excess relatively to the positrons implies a net electric field whose polarization pattern is radial in the shower transverse plane. At first order, the total electric field is oriented following $\mathbf{v}\times\mathbf{B}$ but it is possible to search for second order effect, that can be revealed either by a change in the polarization angle or by an enhanced electric field amplitude due to constructive interferences. Figure~\ref{polscheme} presents the polarization patterns for a vertical shower in a geomagnetic field with a NS component.
In the last year, there have been two different approaches to detect such an additional contribution, compatible with the charge-excess contribution.
\subsection{Evidence for a radial contribution through polarization angle studies}
\begin{center}
\begin{figure}[!ht]
\includegraphics[scale=0.6,draft=false]{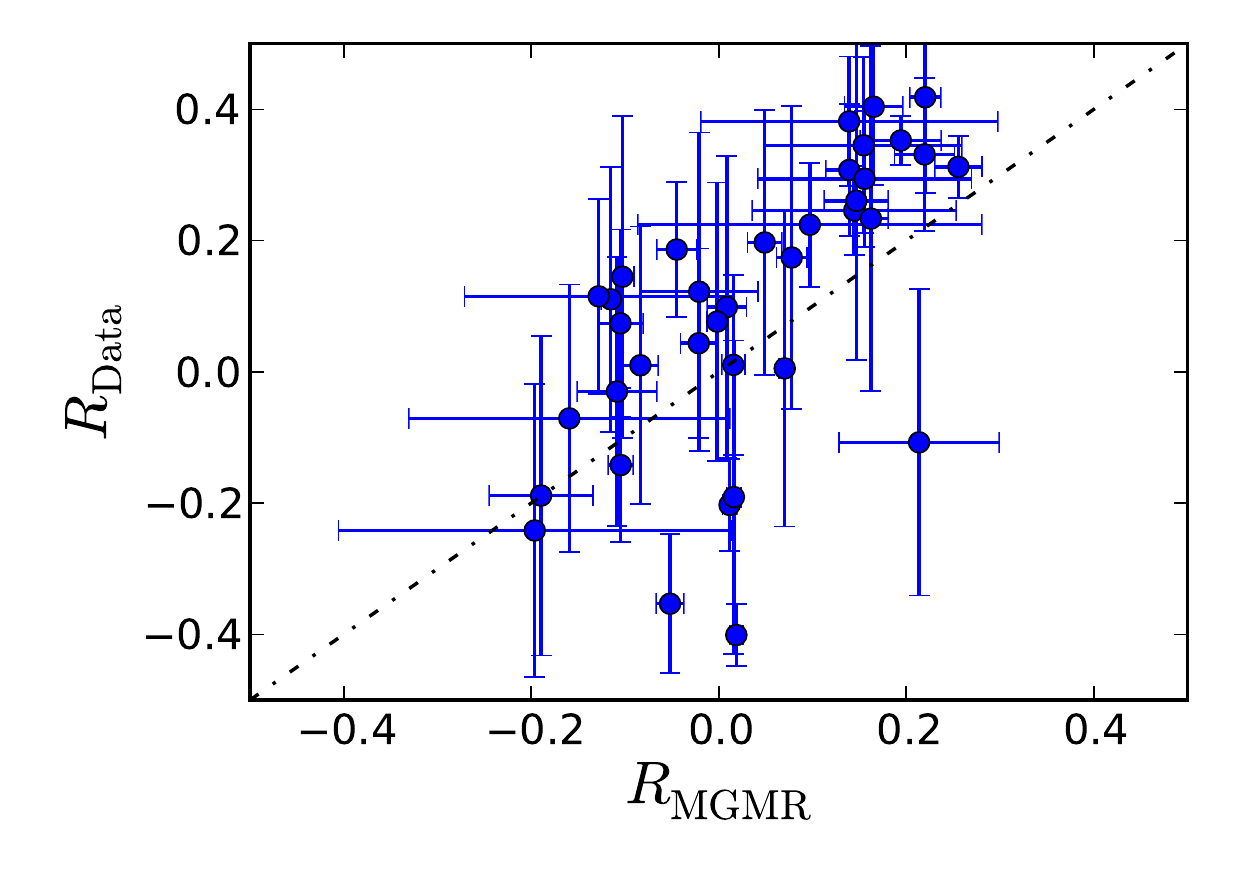}
\caption{Correlation between $R_\mathrm{data}$ and $R_\mathrm{simu}$ for the case of the code MGMR taking into account the charge-excess mechanism in the computation of the simulated electric field. The correlation is clear and much better than for the case where the charge-excess is not included in the simulation code.}\label{cmp01}
\end{figure}
\end{center}

The polarization angle of the total electric field is estimated using the measurements along the EW direction ($x$-axis) and the NS direction ($y$-axis). To detect an electric field contribution incompatible with the geomagnetic mechanism (i.e. having an orientation not aligned along $\mathbf{v}\times\mathbf{B}$), we first construct an observable $R$ characterizing the deviation from a pure geomagnetic electric field, for each radio detector. For that, we first rotate the coordinate system so that the new $x$-axis, $x'$,  is aligned along the direction of $\mathbf{v}\times\mathbf{B}$ projected in the horizontal plane and the new $y$-axis, $y'$, is perpendicular to $x'$. $R$ quantifies the relative signal strength in the direction perpendicular to the geomagnetic expectation. By construction, a pure geomagnetic electric field leads to $R=0$ and an electric field with no geomagnetic component has $R=\pm1$. For a given set of detected showers, we can compute $R_\mathrm{data}$ for the data (we used the data of a pre-AERA prototype) and compute the same factor $R_\mathrm{simu}$ for the same simulated showers (same geometry and detector array). The simulations used here are MGMR~\cite{mgmr2010} and REAS3~\cite{reas2010} and were run using realistic showers (having an excess of electrons) and for showers forced to have the same number of electrons and positrons. The results showed that the correlation is much stronger when including the charge-excess mechanism in the shower. Figure~\ref{cmp01} presents the correlation between $R_\mathrm{data}$ and $R_\mathrm{simu}$ with the charge-excess taken into account for the simulation code MGMR for example.
Quantitatively, the reduced $\chi^2$ decreases from 6.4 (4.7) to 2.7 (3.0) when taking into account the charge-excess in the simulated showers with REAS3 (MGMR).
The conclusion of these results is that the electric field emitted by the showers detected by this pre-AERA prototype is in better agreement with the simulations when the charge-excess mechanism is included in the simulation codes. The relative influence of both prototypes depends strongly on the shower arrival direction and on the relative position of the observer with respect to the shower core. More details on this analysis can be found in~\cite{Abreu:2011ph}. The same analysis is currently performed using the AERA data and more refined simulation codes.

\subsection{Evidence for a radial contribution through core shift}
\begin{center}
\begin{figure}[!ht]
\includegraphics[scale=0.25,draft=false]{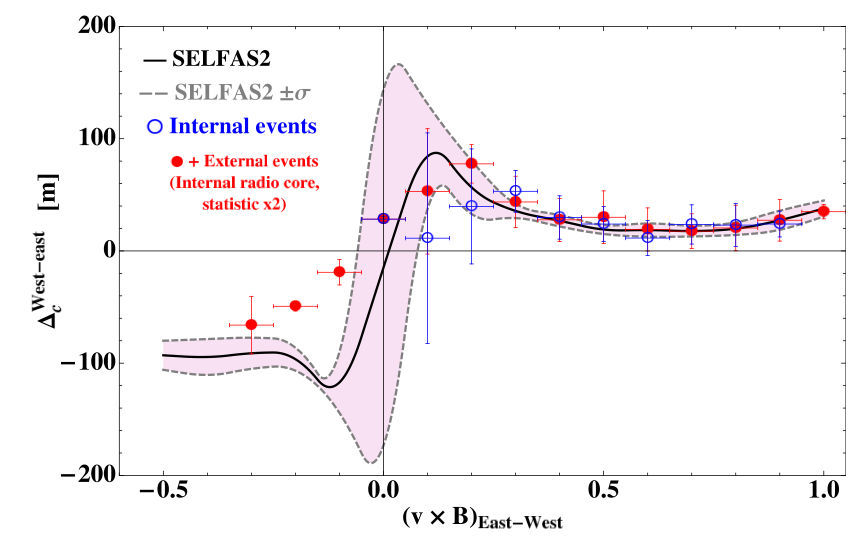}
\caption{Shift between the radio core and the particle core on the EW axis as a function of $(\mathrm{v}\times\mathrm{B})_\mathrm{EW}$. The data are represented by the circles (in blue for internals events with a very reliable reconstruction, in red for external events with a less reliable particle array reconstruction). The shaded zone is the $\pm 1\sigma$ region determined by the simulation with the charge-excess contribution included. No shift is observed on simulated data with no charge-excess.}\label{shift}
\end{figure}
\end{center}

The CODALEMA collaboration reported in~\cite{marinicrc2011} the measurement of a shift toward the east of the shower core position estimated using the radio data (the radio core) with respect to the shower core estimated using the particle array data (the particle core). The particle core position is obtained using the Nishimura-Kamata-Greisen (NKG) lateral distribution (see~\cite{coda2009} for the details). The radio core position is obtained by fitting an exponential function of the type $s(d)=s_0 \exp(-d/d_0)$ where the core coordinates are hidden in the axis distance $d$. The shift to the east is characterized by the quantity $
\Delta_c=x_{c,\mathrm{radio}}-x_{c,\mathrm{particle}}$. This shift is interpreted as the result of the superposition of the geomagnetic and charge-excess electric field components, as presented in Figure~\ref{polscheme}. For this example of a vertical shower, an observer located at the east of the shower core will observe a constructive superposition of the two components of the electric field, contrary to an observer located at the west of the shower core. The radio core will then be reconstructed, in this example, to the east of the actual shower core, defined as the intersection of the shower axis with the ground. The shift strongly depends on the incoming direction through the geomagnetic component. It is therefore interesting to study the shift as a function of the EW component of $\mathrm{v}\times\mathrm{B}$ (because CODALEMA measures only the EW polarization). To check the influence of the charge-excess, simulations were run (using the code SELFAS2~\cite{selfas2011}) with and without the charge-excess mechanism. With no charge-excess, no core shift is observed. Therefore, the core shift is an evidence for the charge-excess contribution. For the selected CODALEMA data set, the dependence of $\Delta_c$ with $(\mathrm{v}\times\mathrm{B})_\mathrm{EW}$ is presented in Figure~\ref{shift} and compared with expectations from simulations with SELFAS2, ran on showers having the same characteristics than the selected showers detected by CODALEMA.

In the last two years, there have been two evidences of a non-geomagnetic contribution to the total electric field. The additional contribution appears to be radially polarized in the transverse shower plane, as it is the case for the charge-excess contribution. More data on a wider energy range will be needed to confirm if this contribution can be associated unambiguously to the charge-excess contribution (Askaryn affect in the air). The AERA experiment should be able to study in great details this contribution.

\section{Sensitivity of the radio signal to the nature of the primary}
The accurate determination shower by shower of the composition of the cosmic rays at ultra-high energies is the next challenge to be taken up. Since some years, using simulations, the radio signal is expected to be correlated to the nature of the primary cosmic ray (see~\cite{Huege200896}, \cite{palmieri2012}). From simulations, it is possible to estimate the value of the atmospheric depth of maximum development of showers $X_\mathrm{max}$, using the LDF of radio profiles. The LOPES collaboration report an uncertainty on $X_\mathrm{max}$ of the order of $150~\mathrm{g~cm}^{-2}$. This large value is dominated by the very noisy environment in Karlsruhe and one could expect (using simulations) a much better resolution ---~around $30~\mathrm{g~cm}^{-2}$~--- in quiet sites. For comparison, the Auger FD has a resolution of $20~\mathrm{g~cm}^{-2}$~\cite{facalicrc2011}. We can expect, by combining data from radio and the SD, to reach a resolution on $X_\mathrm{max}$ close to that of the FD. Another possible method to constrain the nature of the primary has been presented during this conference in~\cite{grebe2012}. The principle of this method is to study the slope (spectral index) of the frequency spectra between 40 and 60~MHz. Simulations show that this spectral index depends not only on the shower geometry but also on the nature of the primary cosmic ray, at a level of 10\% according to MGMR. The dependence on the shower geometry has been confirmed on the AERA data and the possibily to constrain the composition with this method is still under study.

\begin{center}
\begin{figure}[!ht]
\includegraphics[scale=0.35,draft=false]{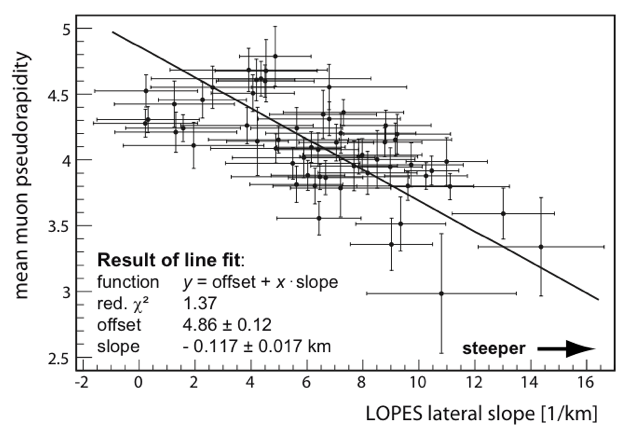}
\caption{Mean muon pseudorapidity as a function of the lateral slope of the 59 selected LOPES radio events.}\label{muonlopes}
\end{figure}
\end{center}

Experimentally, the LOPES collaboration reported~\cite{lopes2012} a first  evidence, at a level of $3.7\sigma$, of the sensitivity of the radio signal to the nature of the primary cosmic ray through the longitudinal profile development. The slope of the radio lateral distribution (using an exponential profile with an estimation of the amplitude at an axis distance of 100~m) is correlated to the mean muon pseudorapidity which is in turn correlated to the shower development: the pseudorapidity is small (large) when the muons are produced at low (high) atmospheric heights. The slope is defined as the factor $a$ in the lateral distribution function: $s(d)=s_{100}\,\exp(-a\,(d-100~\mathrm{m}))$, where $s$ is the electric field amplitude expressed in $\mu\mathrm{V}~\mathrm{m}^{-1}~\mathrm{MHz}^{-1}$ in the band $[40-80]$~MHz and $d$ is the axis distance. Figure~\ref{muonlopes} shows this correlation using a selection of 59 events.

\section{Conclusion}
The most recent and important result of the field is the evidence for a secondary electric field, radially polarized with respect to the shower transverse plane. This electric field could well be the contribution of electrons in excess in the shower, known as the Askaryan effect in the air. The confirmation of such observation and its quantification would be a major advance in the understanding of the emission processes of electric field in air showers. The radio signal, supposed to be correlated to the longitudinal profile of the shower, is very promising in the estimation of the composition of ultra-high energy cosmic rays. The AERA experiment, located at the north-west of the Auger SD, it installed at an ideal site because showers can be detected by many detectors: the Infill Auger SD, the regular Auger SD, the regular Auger FD, AMIGA (dedicated to the muonic contents of the shower,~\cite{Abreu:2011ph}) and HEAT~\cite{Abreu:2011ph} which permits to detect the fluorescence light at higher elevation angles corresponding to lower energy showers. All these detectors will study in great details the showers in the energy range $10^{17}$ to $10^{18.5}$, providing high-quality data to study the transition from a galactic to an extra-galactic origin of cosmic rays. The phase 2 of AERA, providing a total of 161 stations spread over 20~km$^2$ is starting in March 2013.





\bibliographystyle{aipproc}   

\bibliography{biblio,brevenu_proceedings_speaker,brevenu_proceedings,brevenu_spires}

\IfFileExists{\jobname.bbl}{}
 {\typeout{}
  \typeout{******************************************}
  \typeout{** Please run "bibtex \jobname" to optain}
  \typeout{** the bibliography and then re-run LaTeX}
  \typeout{** twice to fix the references!}
  \typeout{******************************************}
  \typeout{}
 }

\end{document}